\documentclass[runningheads]{llncs}
\usepackage{graphicx}
\usepackage{amsmath}
\usepackage{amssymb}
\usepackage{caption}
\usepackage{diagbox}
\usepackage{subfigure}
\usepackage{mathtools}
\usepackage{stmaryrd}
\usepackage[mathscr]{eucal}
\usepackage[linesnumbered,lined,boxed,commentsnumbered]{algorithm2e}
\usepackage{comment}
\usepackage{floatrow}
\usepackage{multirow}

\usepackage[rightcaption]{sidecap}
\usepackage{soul}

\usepackage{hyperref}
\hypersetup{
    colorlinks=true,
    linkcolor=blue,
    filecolor=magenta,      
    urlcolor=cyan,
}

\begin{document}

\title{Weakly Supervised Registration of Prostate MRI and Histopathology Images}
\author{Wei Shao\inst{1}, Indrani Bhattacharya\inst{1}, Simon J.C. Soerensen\inst{2}, Christian A. Kunder\inst{3}, Jeffrey B. Wang\inst{4}, Richard E. Fan\inst{2},  Pejman Ghanouni\inst{1}, James D. Brooks\inst{2}, Geoffrey A. Sonn\inst{1,2}, Mirabela Rusu\inst{1}}
\institute{Department of Radiology, Stanford University, Stanford, CA 94305, USA\\ 
\and Department of Urology, Stanford University, Stanford, CA 94305, USA\\ 
\and Department of Pathology, Stanford University, Stanford, CA 94305, USA \\
\and School of Medicine, Stanford University, Stanford, CA 94305, USA}
\maketitle
\begin{abstract}
The interpretation of prostate MRI suffers from low agreement across radiologists due to the subtle differences between cancer and normal tissue.
Image registration addresses this issue by accurately mapping the ground-truth cancer labels from surgical histopathology images onto MRI.
Cancer labels achieved by image registration can be used to improve radiologists' interpretation of MRI by training deep learning models for early detection of prostate cancer.
A major limitation of current automated registration approaches is that they require manual prostate segmentations, which is a time-consuming task, prone to errors. 
This paper presents a weakly supervised approach for affine and deformable registration of MRI and histopathology images without requiring prostate segmentations.
We used manual prostate segmentations and mono-modal synthetic image pairs to train our registration networks to align prostate boundaries and local prostate features.
Although prostate segmentations were used during the training of the network, such segmentations were not needed when registering unseen images at inference time.
We trained and validated our registration network with 135 and 10 patients from an internal cohort, respectively.
We tested the performance of our method using 16 patients from the internal cohort and 22 patients from an external cohort.
The results show that our weakly supervised method has achieved significantly higher registration accuracy than a state-of-the-art method run without prostate segmentations.
Our deep learning framework will ease the registration of MRI and histopathology images by obviating the need for prostate segmentations.
\end{abstract}

\section{Introduction}
Prostate cancer is the most frequently diagnosed solid-organ cancer and the second leading cause of cancer death among American men~\cite{prostate_cancer_stats}.
Contemporary prostate cancer screening and diagnosis is marred by underdetection of aggressive cancers and overdiagnosis and overtreatment of low-risk prostate cancers.
Magnetic resonance imaging (MRI) is increasingly used to improve prostate cancer diagnosis in patients with elevated PSA levels~\cite{TurkbeyL.Baris2012MMap}.
However, the interpretation of MRI suffers from low inter-reader agreement across radiologists coupled with large variations in reported sensitivity (58-96\%) and specificity (23-87\%)~\cite{AhmedHashimU2017Daom}.
This is largely due to the lack of histologic confirmation of the extent of cancer areas on MRI. 
One solution to this challenge is to map the ground-truth extent of prostate cancer from surgical histopathology images onto pre-operative MRI using image registration.
Radiologists can use such mapping as a training tool to learn from subjects that have already undergone radical prostatectomy.
Moreover, accurate cancer labels on MRI can also be used to develop and validate machine learning approaches for early detection of prostate cancer on MRI~\cite{correlation,bhattacharya2020corrsignet,seetharaman2021automated}.

Previous approaches require manual prostate segmentations to facilitate the challenging MRI-histopathology registration~\cite{ChappelowJonathan2011Erom,ReynoldsH.M.2015Doar,WuHoldenH.2019Asup,rusu2020registration,shao2021prosregnet,sood20213d}.
While prostate segmentations on MRI are required for ultrasound-MRI fusion targeted biopsy, such segmentations are not routinely performed in men undergoing prostatectomy after conventional prostate biopsy. Moreover, pathologists do not segment the prostate on digital histopathology. Since prostate segmentations on MRI and histopathology are not always immediately available and gland segmentation is a time-consuming task prone to errors especially at the base and apex of the prostate, there is value in registration of MRI and histopathology without the need for gland segmentation.

Here, we present a weakly supervised MRI-histopathology registration approach that does not require prostate segmentations during the testing.
The major advantage of our method is that it avoids the above shortcomings of prostate segmentation by not relying on it at inference. 
The direct registration of prostate MRI and histopathology images without using prostate segmentations is a challenging task for two reasons.
First, the appearance of the prostate in the two modalities differs substantially including intensity characteristics and image resolutions.
Second, the size and the location of the prostate on the MRI and histopathology images varies across patients.
To address the above challenges, we trained our registration neural network with manual prostate segmentations and mono-modal synthetic image pairs so that the network is able to align both prostate boundaries and local prostate features.
We acknowledge that many weakly supervised registration methods have been proposed for image registration \cite{hu2018label,BalakrishnanGuha2019VALF,deVosBobD2019Adlf}. Our method used both label-driven and image-driven losses, unlike Hu et al. that only used label-based loss \cite{hu2018label}. 
Moreover, others used multimodal losses for the training \cite{BalakrishnanGuha2019VALF,deVosBobD2019Adlf}, however, these methods are not suitable for registration of MRI and histopathology due to their large differences in appearance. 
The paper has the following three major contributions.
\begin{itemize}
	\item Our MRI-histopathology registration neural network does not require prostate segmentations during the testing.
	\item Our registration neural network was trained with mono-modal synthetic image pairs and manual prostate segmentations so that it can be used for affine and deformable registration of multi-modal images.
	\item An unsupervised intensity loss, an auxiliary segmentation loss, and  a regularization loss were developed for the challenging registration problem.
\end{itemize}

\section{Methods}
\subsection{Data Acquisition}
This institutional review board-approved study included 183 patients with biopsy-confirmed prostate cancer from two cohorts.  
The first cohort is an internal cohort consisting of 161 patients with a pre-operative T2-weighted (T2-w) MRI and whole-mount digital histopathology images from radical prostatectomy.
Slice-to-slice correspondences between the histopathology images and the T2-w MRI were optimized using customized 3D-printed molds.
The second cohort is an external validation cohort consisting of 22 patients from the publicly available TCIA "Prostate-MRI" dataset~\cite{ChoykePeter2016DFP}.
Each patient had a pre-operative T2-w MRI and whole-mount histopathology images from radical prostatectomy. 
3D-printed molds were also used to section the prostate specimen in the same plane as the MRI. 
An intensity standardization algorithm~\cite{nyul2000new} was used to correct for misaligned MR image intensities between patients.
For both cohorts, the prostate, urethra, and anatomic landmarks on MRI and histopathology images were manually segmented.
Cancerous regions on the histopathology images were also manually segmented.

\subsection{Overview of Proposed Method}
The goal of an image registration network is to find a bijective and smooth geometric transformation $\phi$ that defines the point-to-point correspondences between a moving image $I_m$ and a fixed image $I_f$.
\begin{figure}[!hbt]
	\centering
        \includegraphics[width=\linewidth]{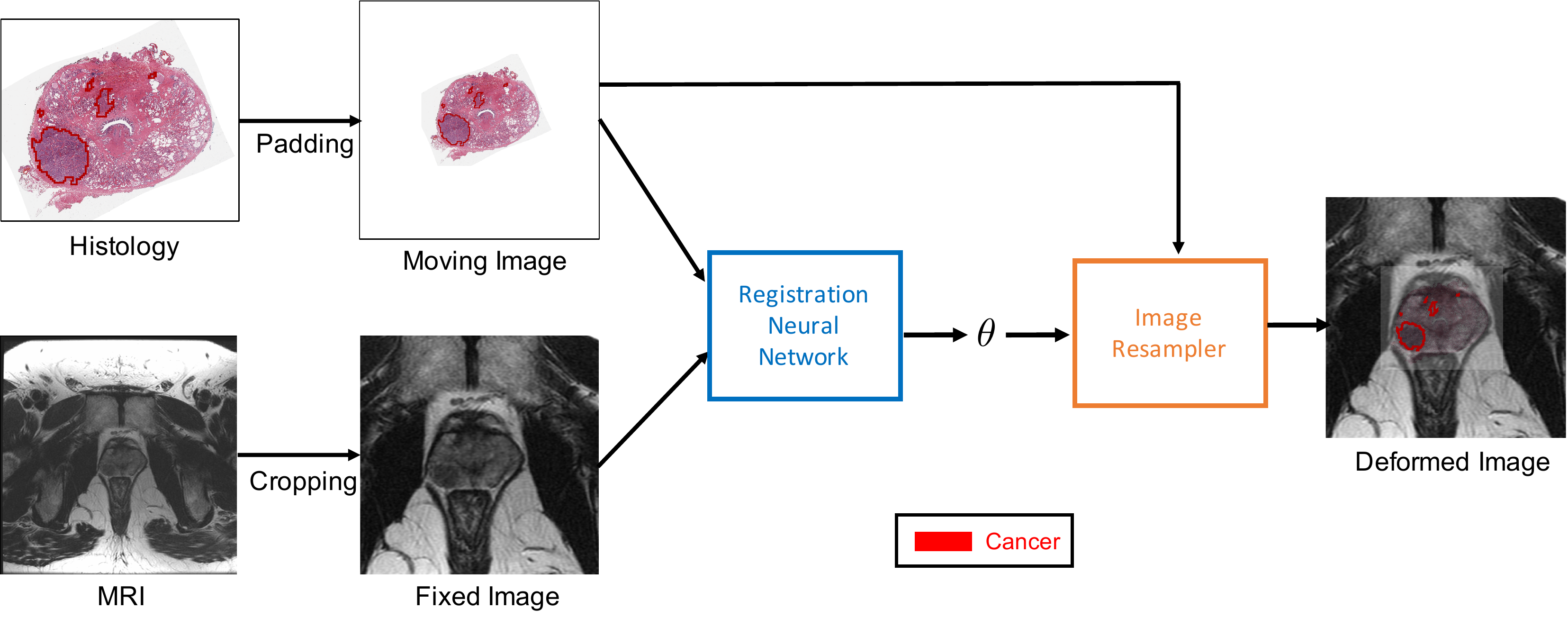}
	\caption{
	An overview of our weakly supervised learning registration pipeline. 
		\label{fig:overview}}
\end{figure}
Figure~\ref{fig:overview} presents an overview of our weakly supervised registration pipeline.
The input to our registration pipeline is a histopathology image and the corresponding axial MRI slice.
First, we cropped a 100mm$\times$100mm central region from the MRI slice as the fixed image.
The cropped MRI is sufficiently large to cover the entire prostate since the prostate is located approximately in the center of the MRI and the average size of the prostate in the axial plane is 3mm$\times$4mm.
We padded the histopathology image to the same size as the cropped MRI, i.e., 100mm$\times$100mm.
The histopathology image served as the moving image.
The registration neural network then takes the fixed image the moving image as the input and outputs a vector $\theta$ that parameterizes a composite transformation (affine + deformable) between the two images.
Finally, the predicted transformation is used to deform the ground-truth cancer labels from the histopathology image onto the MRI.

\subsection{Registration Neural Network}
\begin{figure}[!hbt]
	\centering
        \includegraphics[width=\linewidth]{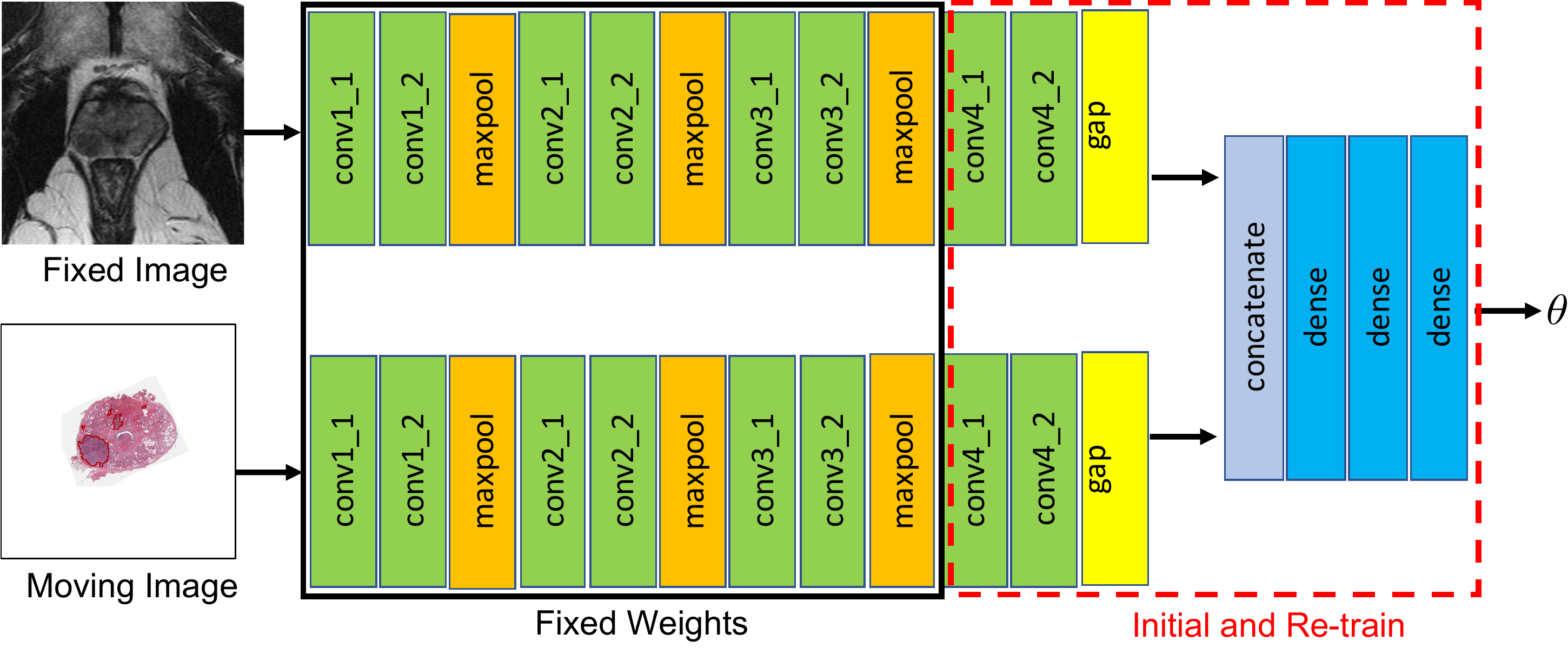}
	\caption{
Proposed registration neural network. Fixed Weights correspond to layers whose weights were frozen during the training, and Initial and Re-train correspond to layers whose weights were updated during the training.
		\label{fig:regnet}}
\end{figure}

Our registration neural network shown in Fig.~\ref{fig:regnet} is considered as a function that maps a fixed image and a moving image to a vector $\theta$ that parameterizes a geometric transformation between the two images.
Features in the fixed and moving images were extracted  by the first few layers of the VGG16 network~\cite{SimonyanKaren2015VDCN} pre-trained on the ImageNet dataset~\cite{deng2009imagenet}.
During the training, we freeze all layers of the VGG16 network except for the last two convolutional layers.
Inspired by~\cite{deVosBobD2019Adlf}, the extracted image features were connected to the global average pooling (GAP) layer to transform their dimensions from $(N\times N \times C)$ to $(1\times 1 \times C)$, where $N$ is the size of a feature image and $C$ is the number of feature images.
We concatenated the outputs of the GAP layers into a column vector, which was connected to a stack of three dense layers to regress the parameter vector $\theta$.
One advantage of using the GAP layer is that it can convert feature images of any dimension to $1\times 1$, and thus it allows our neural network to have fixed and moving images of different sizes.
The GAP layer could also prevent the neural network from overfitting since it has reduced the complexity the model.

\subsection{Transformation Model}
For each pair of images, our registration network predicts an affine transformation followed by a nonrigid transformation.
To parameterize a 2D affine transformation, we set the length of the output parameter vector $\theta$ from the registration network to 6.
Instead of directly using $\theta$ as the affine matrix, we define the affine transformation $\phi$ parameterized by $\theta$ as
\begin{equation}
    \phi(x,y) = \begin{bmatrix}
    1 + \alpha \theta_1 & \alpha \theta_2 \\ \alpha \theta_4 & 1 + \alpha \theta_5
    \end{bmatrix} \begin{bmatrix} x \\y \end{bmatrix} + \begin{bmatrix} \alpha \theta_3 \\ \alpha \theta_6 \end{bmatrix}
\end{equation}
where $(x,y)$ is a pixel in the fixed image $I_f$, and $\alpha = 0.001$ is a constant.
The purpose of scaling $\theta$ by a small constant and adding an identity matrix is to guarantee that the transformation $\phi$ is  close to the identity at the beginning of the training, which would substantially improve the robustness of our model.

In this paper, we parameterize nonrigid transformations by a $4\times 4$ thin-plate spline grid~\cite{donato2002approximate}.
Therefore, the length of the parameter vector $\theta$ is set to $2\times 4 \times 4 = 32$.
Similarly, we scale the parameter vector $\theta$ by a small constant $\alpha = 0.001$ and add an identity matrix so that the initial estimation of the nonrigid transformation $\phi$ is close to the identity transformation.

\subsection{Loss Functions}
In this paper, we propose three loss functions: an unsupervised intensity loss $L_{int}$ computed on two synthetic mono-modal image pairs, an auxiliary segmentation loss $L_{seg}$ computed on two prostate segmentations, and a regularization loss $L_{reg}$ that measures the smoothness of nonrigid transformations.

Our registration framework consists of  an affine registration network and a deformable registration network.
We trained the two registration networks simultaneous and used the output of the affine network as an initialization of the deformable network.
The loss function of the affine registration network is defined as
\begin{equation}
    L_{affine} = L_{seg} + 0.05 L_{int}
\end{equation}

The loss function of the deformable registration network is defined as
\begin{equation}
    L_{def} = L_{seg} + 0.05 L_{int} + 0.05 L_{reg}
\end{equation}

Each training example $(I_f, I_m, S_f, S_m)$ consists of four images, where $I_f$ and $I_m$ are the fixed and moving images, and $S_f$ and $S_m$ are the corresponding prostate segmentations.
We applied two random transformations to deform $I_f$ and $I_m$ into $I_f'$ and $I_m'$.
The mono-modal image pairs $(I_f, I_f')$ and $(I_m, I_m')$ were fed into the registration network to predict two transformations $\phi_{mono}^f$ and $\phi_{mono}^m$.
We define the unsupervised intensity loss $L_{int}$ as
\begin{equation}
    L_{int} = MSE(I_f, I_f'(\phi_{mono}^f)) + MSE(I_m, I_m'(\phi_{mono}^m))
\end{equation}
where $MSE(\cdot,\cdot)$ is the mean squared error. The motivation of using $L_{int}$ is to train the network to align local prostate features irrespective of image modalities. 

We used manual prostate segmentations $S_f$ and $S_m$ to train the network to align prostate boundaries.
Let $\phi_{multi}$ denote the transformation between $I_f$ and $I_m$ predicted by the registration neural network.
We define the auxiliary segmentation loss $L_{seg}$ as
\begin{equation}
L_{seg} = Dice(S_f, S_m(\phi_{multi})) =  \frac{2|S_f \cap S_m(\phi_{multi})|}{|S_f| + |S_m(\phi_{multi})|}
\label{eq:dice}
\end{equation}
where $|\cdot|$ is cardinality of a set.

We used the regularization loss $L_{reg}$ to measure the smoothness of thin-plate spline transformations $\phi$
\begin{equation}
    L_{reg} = ||Lu||^2_{L^2}
\end{equation}
where $u = \phi - Id$, $L = -0.75 \nabla^2 - 0.25 \nabla (\nabla \cdot) + 0.01 I$, $\nabla$ is the gradient operator, $\nabla \cdot$ is the divergence operator, and $I$ is the identity matrix. 

\subsection{Previous Method}
The state-of-the-art ProsRegNet method~\cite{shao2021prosregnet} is the only approach using deep learning for the registration of MRI and histopathology and was used to compare with our weakly supervised method.
We downloaded the code from the public repository \url{https://github.com/pimed//ProsRegNet}.
We acknowledge the following differences between the proposed method and the ProsRegNet method.
\begin{itemize}
\item Network architecture. The ProsRegNet method used ResNet101 while our method used VGG16 for feature extraction. The ProsRegNet method used a correlation layer to compute the correlation between feature maps while our method used the global average pooling layer to reduce the size of each feature map. 
\item Loss function. The ProsRegNet method used the mean squared error (MSE) as the training loss. Our proposed method used MSE and two additional terms: the Dice coefficient loss to align the prostate boundaries and the regularization loss to guarantee the smoothness of the transformations.
\item The ProsRegNet method used masked synthetic unimodal image pairs for the training. The proposed method used unmasked synthetic unimodal image pairs, unmasked real multimodal image pairs, and manual prostate masks for the training.   
\end{itemize}

\subsection{Evaluation Metrics} 
We use Dice coefficient defined in Eq.~\ref{eq:dice} to measures the overlap between prostate segmentation of the fixed image ($S_f$) and prostate segmentation of the deformed image ($S_m(\phi)$).

We use Hausdorff distance to measure the distance between boundaries of the prostate segmentations
\begin{equation}
Haus = \max\left\{ \sup_{a\in S_f} \inf_{b\in S_m(\phi)} ||a - b||,\sup_{b\in S_m(\phi)} \inf_{a\in S_f}||a - b||\right\}.
\end{equation}

We use the mean landmark error (MLE) to measure the accuracy of estimated point-to-point correspondences 
\begin{equation}
MLE = \frac{1}{N}\sum_{i=1}^{N}||p'_i - \phi(p_i)||
\label{eq:lmk_error}
\end{equation}
where $N$ is the number of landmarks, $p_1,\cdots,p_N$ and $p'_1,\cdots,p'_N$ are landmarks in the fixed image and moving images, respectively.

We also use Eq.~\ref{eq:lmk_error} to evaluate the distance between the center of urethra segmented on the fixed image and the deformed image, i.e., urethra deviation.

\subsection{Experimental Design}
We trained and validated our registration networks with 135 and 10 patients from the internal cohort, respectively.
We used an initial learning rate of 0.001, a learning rate decay of 0.9, a batch size of 1, 50 epochs, and the Adam optimizer~\cite{KingmaDiederik2017AAMf}.
We tested the performance of our method using 16 patients from the internal cohort and 22 patients from the external cohort.
We compared our weakly supervised registration approach to the previous ProsRegNet deep learning registration approach~\cite{shao2021prosregnet}.
Since the goal of this paper is to develop a registration approach that does not require prostate segmentation, both our method and the competing ProsRegNet method were tested without using prostate segmentations.
All experiments were performed on the NVIDIA Quadro GV100 GPU (32GB memory, 5120 CUDA cores, 640 tensor cores).

\section{Results}
\subsection{Qualitative Results}
\begin{figure}[!htb] 
	\centering
        \includegraphics[width=0.8\linewidth]{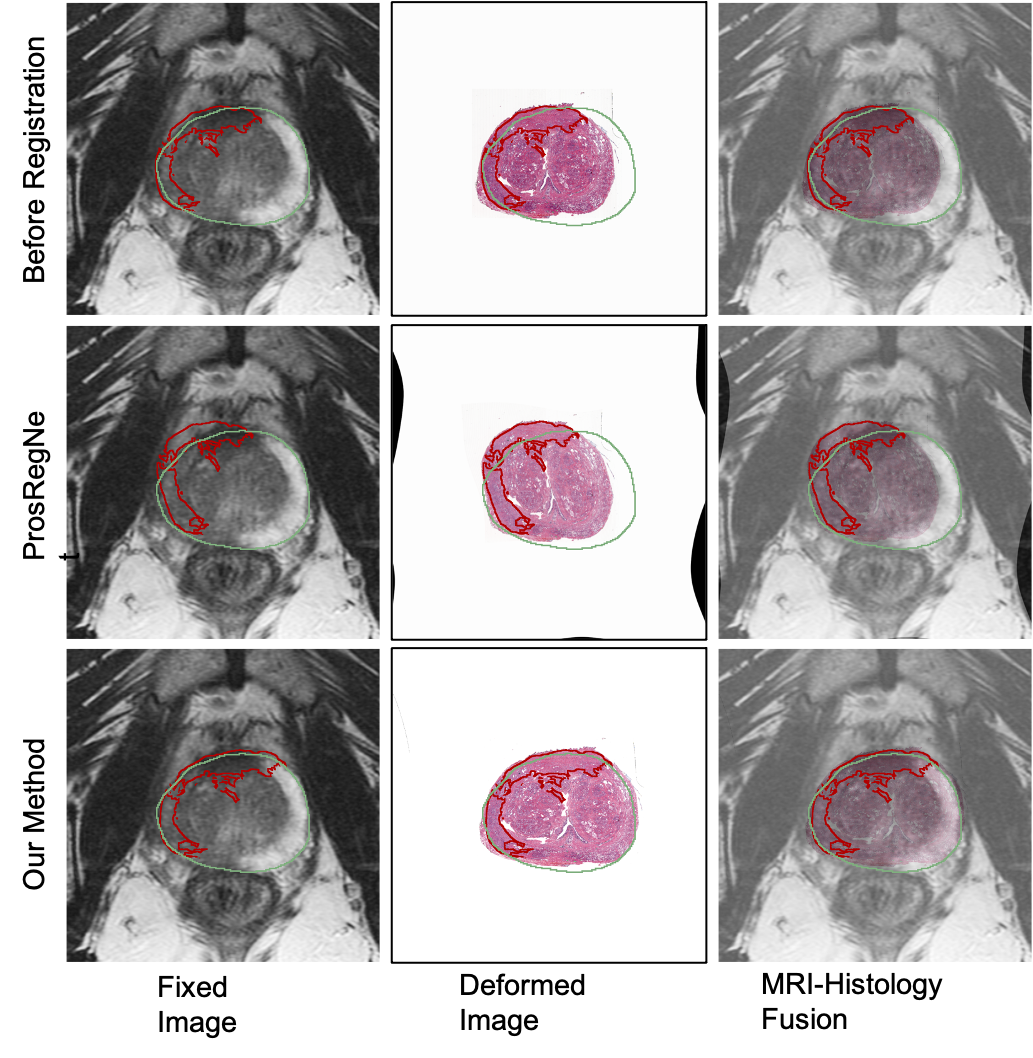}
	\caption{
Comparison of the registration results of our weakly supervised method and previous ProsRegNet method of a representative subject from the internal cohort. The green outlines correspond to manual segmentation of the prostate on the fixed MRI image. The red outlines correspond to the boundaries of the cancer labels which were obtained by deforming the ground-truth cancer labels from the histopathology images onto MRI using the resulting transformations from image registration.
		\label{fig:reg-results}}
\end{figure}

In Figure~\ref{fig:reg-results}, we compared the registration results of our weakly supervised method and the ProsRegNet method of a representative patient from the internal cohort.
The first row of Fig.~\ref{fig:reg-results} shows that there is large misalignment between the MRI and histopathology images before image registration.
The second row of Fig.~\ref{fig:reg-results} shows that the previous ProsRegNet barely improved the registration between the two images.
The third row of Fig.~\ref{fig:reg-results} shows that our weakly supervised method has not only accurately aligned prostate boundaries, but also local features inside the prostate which includes the urethra, benign prostatic hyperplasia regions, and cancerous regions.
This example demonstrates the  superiority of our weakly supervised method over the previous ProsRegNet method.
Accurate transformations predicted our weakly supervised registration network enables accurate mapping of the ground-truth cancer labels from the histopathology image onto MRI.

\subsection{Quantitative Results}
Table~\ref{table:stats} summarizes the average Dice coefficients, Hausdorff distances, urethra deviations, and landmark errors before and after our weakly supervised registration and the ProsRegNet registration.
The Dice coefficients and Hausdorff distances show that our weakly supervised approach has achieved significantly more accurate alignment of the prostate boundaries for both cohorts ($p < 0.05$).
The urethra deviations and landmark errors show that our weakly supervised approach has also achieved much more accurate alignment of local prostate features than the ProsRegNet approach for both cohorts.
One major reason for the poor performance of the ProsRegNet model is that it was trained using images masked by prostate segmentations while prostate segmentations were not provided during the testing.
\begin{table*}[!hbt]
	\caption{Registration results before and after our weakly supervised registration and the ProsRegNet registration of two cohorts. Cohort 1 is our internal cohort and Cohort 2 is an external cohort.}
	\begin{center}
		\begin{tabular}{|c|c|c|c|c|c|}
			\hline	\multirow{2}{*}{Dataset} & Registration	& Dice & Hausdorff& Urethra & Landmark \\
			& Approach & Coefficient  & Distance & Deviation & Error \\
			[0.5ex] 
			\hline
			\hline	\multirow{3}{*}{Cohort 1}& Input & 0.80 ($\pm$ 0.06) & 7.57 ($\pm$ 2.21) & 4.32 ($\pm$ 2.29)& 4.48 ($\pm$ 1.77)\\
			& ProsRegNet & 0.80 ($\pm$ 0.06) & 7.76 ($\pm$ 2.00) & 4.54 ($\pm$ 2.46)& 4.58 ($\pm$ 1.87) \\
			&Our Method & \textbf{0.90 ($\pm$ 0.02)} & \textbf{4.38 ($\pm$ 0.64)} & \textbf{3.09 ($\pm$ 1.06)}& \textbf{3.09 ($\pm$ 0.81)}\\
			\hline
			\hline	\multirow{3}{*}{Cohort 2} & Input & 0.83 ($\pm$ 0.05) & 7.23 ($\pm$ 2.30) & 3.39 ($\pm$ 1.47)& 3.40 ($\pm$ 1.04)\\
			& ProsRegNet & 0.82 ($\pm$ 0.04) & 8.00 ($\pm$ 2.12) & 3.40 ($\pm$ 1.34) & 4.09 ($\pm$ 1.38)\\
			& Our Method  & \textbf{0.89 ($\pm$ 0.03)} & \textbf{5.01 ($\pm$ 0.97)} & \textbf{2.59 ($\pm$ 1.16)}& \textbf{3.13 ($\pm$ 0.76)}\\
			\hline
			\hline
		\end{tabular}
	\end{center}
	\label{table:stats}
\end{table*}

\section{Discussion and Conclusion}
This paper presents a weakly supervised learning approach for the affine and deformable image registration of MRI and histopathology images.
Unlike previous fully-automated  approaches,  our  approach  does  not  require  manual  segmentations of the prostate during the testing. 
Therefore, our approach will ease the registration of MRI and histopathology images by obviating the need for prostate segmentations.
Moreover,  previous  registration  approaches are  sensitive  to errors in prostate  segmentations.
Therefore, fine-tuning of the prostate segmentations is a necessity for previous approaches to achieve accurate registration results. 
Although our model was trained with patients from an internal cohort, the results show that the trained model can be generalized to accurately register unseen images from an external cohort. 
We acknowledge that the use of patients undergoing prostatectomy might introduce spectrum bias in our cohort that does not translate into other populations (e.g., candidates for active surveillance). However, we used grade as a surrogate for the aggressive disease since it has been demonstrated to be the most powerful predictor of outcome in localized prostate cancer. Moreover, grade is also used for clinical decisions (e.g., selection of patients for active surveillance). As such, radical prostatectomy cases are ideal for model building as they allow: (1) precise registration of MR and histopathology images, (2) provide ground truth cancer label at every voxel within the prostate and (3) diversity of low- and high-grade cancers as they often coexist within the same lesion. However, such granularity of labels cannot be achieved in subjects that only undergo biopsy as the prostate is only sampled at the biopsy sites. 
The proposed algorithm can relieve radiologists the task of manual prostate segmentation while generating accurate cancer labels on MRI. 
Accurate labels can facilitate clinical care and facilitate use of deep learning approaches for detection of aggressive cancers on preoperative MRI.

\bibliographystyle{unsrt}
\bibliography{main} 

\begin{thebibliography}{10}

\bibitem{prostate_cancer_stats}
{American Cancer Society}.
\newblock {\em {Facts \& Figures 2020}}.
\newblock American Cancer Society, Atlanta, GA, 2021.

\bibitem{TurkbeyL.Baris2012MMap}
Baris Turkbey, L. and Peter Choyke, L.
\newblock {Multiparametric MRI and prostate cancer diagnosis and risk
  stratification}.
\newblock {\em Current Opinion in Urology}, 22(4):310--315, 2012.

\bibitem{AhmedHashimU2017Daom}
Hashim~U Ahmed, Ahmed El-Shater~Bosaily, Louise~C Brown, Rhian Gabe, Richard
  Kaplan, Mahesh~K Parmar, et~al.
\newblock Diagnostic accuracy of multi-parametric mri and trus biopsy in
  prostate cancer (promis): a paired validating confirmatory study.
\newblock {\em The Lancet}, 389(10071):815--822, 2017.

\bibitem{correlation}
Catherine~Elizabeth Lovegrove, Mudit Matanhelia, Jagpal Randeva, David
  Eldred-Evans, Henry Tam~Saiful Miah, Mathias Winkler, et~al.
\newblock {The Role of Pathology Correlation Approach in Prostate Cancer Index
  Lesion Detection and Quantitative Analysis with Multiparametric MRI}.
\newblock {\em NIH}, 2016.

\bibitem{bhattacharya2020corrsignet}
Indrani Bhattacharya, Arun Seetharaman, Wei Shao, Rewa Sood, Christian~A
  Kunder, Richard~E Fan, Simon John~Christoph Soerensen, Jeffrey~B Wang, Pejman
  Ghanouni, Nikola~C Teslovich, et~al.
\newblock Corrsignet: Learning correlated prostate cancer signatures from
  radiology and pathology images for improved computer aided diagnosis.
\newblock In {\em International Conference on Medical Image Computing and
  Computer-Assisted Intervention}, pages 315--325. Springer, 2020.

\bibitem{seetharaman2021automated}
Arun Seetharaman, Indrani Bhattacharya, Leo~C Chen, Christian~A Kunder, Wei
  Shao, Simon~JC Soerensen, Jeffrey~B Wang, Nikola~C Teslovich, Richard~E Fan,
  Pejman Ghanouni, et~al.
\newblock Automated detection of aggressive and indolent prostate cancer on
  magnetic resonance imaging.
\newblock {\em Medical Physics}, 2021.

\bibitem{ChappelowJonathan2011Erom}
Jonathan Chappelow, B.~Nicolas Bloch, Neil Rofsky, Elizabeth Genega, Robert
  Lenkinski, William Dewolf, et~al.
\newblock {Elastic registration of multimodal prostate MRI and histology via
  multiattribute combined mutual information}.
\newblock {\em Medical Physics}, 38(4):2005--2018, 2011.

\bibitem{ReynoldsH.M.2015Doar}
H.~M. Reynolds, S.~Williams, A.~Zhang, R.~Chakravorty, D.~Rawlinson, C.~S. Ong,
  et~al.
\newblock {Development of a registration framework to validate MRI with
  histology for prostate focal therapy}.
\newblock {\em Medical Physics}, 42(12):7078--7089, 2015.

\bibitem{WuHoldenH.2019Asup}
Holden~H. Wu, Alan Priester, Pooria Khoshnoodi, Zhaohuan Zhang, Sepideh
  Shakeri, Sohrab Afshari~Mirak, et~al.
\newblock {A system using patient-specific 3D-printed molds to spatially align
  in vivo MRI with ex vivo MRI and whole-mount histopathology for prostate
  cancer research}.
\newblock {\em Journal of Magnetic Resonance Imaging}, 49(1), 2019.

\bibitem{rusu2020registration}
Mirabela Rusu, Wei Shao, Christian~A Kunder, Jeffrey~B Wang, Simon~JC
  Soerensen, Nikola~C Teslovich, Rewa~R Sood, Leo~C Chen, Richard~E Fan, Pejman
  Ghanouni, et~al.
\newblock Registration of presurgical mri and histopathology images from
  radical prostatectomy via rapsodi.
\newblock {\em Medical physics}, 47(9):4177--4188, 2020.

\bibitem{shao2021prosregnet}
Wei Shao, Linda Banh, Christian~A Kunder, Richard~E Fan, Simon~JC Soerensen,
  Jeffrey~B Wang, Nikola~C Teslovich, Nikhil Madhuripan, Anugayathri Jawahar,
  Pejman Ghanouni, et~al.
\newblock Prosregnet: A deep learning framework for registration of mri and
  histopathology images of the prostate.
\newblock {\em Medical image analysis}, 68:101919, 2021.

\bibitem{sood20213d}
Rewa~R Sood, Wei Shao, Christian Kunder, Nikola~C Teslovich, Jeffrey~B Wang,
  Simon~JC Soerensen, Nikhil Madhuripan, Anugayathri Jawahar, James~D Brooks,
  Pejman Ghanouni, et~al.
\newblock 3d registration of pre-surgical prostate mri and histopathology
  images via super-resolution volume reconstruction.
\newblock {\em Medical Image Analysis}, 69:101957, 2021.

\bibitem{hu2018label}
Yipeng Hu, Marc Modat, Eli Gibson, Nooshin Ghavami, Ester Bonmati, Caroline~M
  Moore, Mark Emberton, J~Alison Noble, Dean~C Barratt, and Tom Vercauteren.
\newblock Label-driven weakly-supervised learning for multimodal deformable
  image registration.
\newblock In {\em 2018 IEEE 15th International Symposium on Biomedical Imaging
  (ISBI 2018)}, pages 1070--1074. IEEE, 2018.

\bibitem{BalakrishnanGuha2019VALF}
Guha Balakrishnan, Amy Zhao, Mert~R Sabuncu, John Guttag, and Adrian~V Dalca.
\newblock Voxelmorph: A learning framework for deformable medical image
  registration.
\newblock {\em IEEE Transactions on Medical Imaging}, 38(8):1788--1800, August
  2019.

\bibitem{deVosBobD2019Adlf}
Bob~D de~Vos, Floris~F Berendsen, Max~A Viergever, Hessam Sokooti, Marius
  Staring, and Ivana IÅ¡gum.
\newblock A deep learning framework for unsupervised affine and deformable
  image registration.
\newblock {\em Medical Image Analysis}, 52:128--143, 2019.

\bibitem{ChoykePeter2016DFP}
Peter Choyke, Baris Turkbey, Peter Pinto, Maria Merino, and Brad Wood.
\newblock Data from prostate-mri, 2016.

\bibitem{nyul2000new}
L{\'a}szl{\'o}~G Ny{\'u}l, Jayaram~K Udupa, and Xuan Zhang.
\newblock New variants of a method of mri scale standardization.
\newblock {\em IEEE transactions on medical imaging}, 19(2):143--150, 2000.

\bibitem{SimonyanKaren2015VDCN}
Karen Simonyan and Andrew Zisserman.
\newblock Very deep convolutional networks for large-scale image recognition.
\newblock {\em arXiv.org}, 2015.

\bibitem{deng2009imagenet}
Jia Deng, Wei Dong, Richard Socher, Li-Jia Li, Kai Li, and Li~Fei-Fei.
\newblock Imagenet: A large-scale hierarchical image database.
\newblock In {\em 2009 IEEE conference on computer vision and pattern
  recognition}, pages 248--255. Ieee, 2009.

\bibitem{donato2002approximate}
Gianluca Donato and Serge Belongie.
\newblock Approximate thin plate spline mappings.
\newblock In {\em European conference on computer vision}, pages 21--31.
  Springer, 2002.

\bibitem{KingmaDiederik2017AAMf}
Diederik Kingma and Jimmy Ba.
\newblock Adam: A method for stochastic optimization.
\newblock {\em arXiv.org}, 2017.

\end{thebibliography}
\end{document}